
\magnification = 1200

\def\gboxit#1{\hbox{\vrule\vbox{\hrule\kern3pt\vtop
{\hbox{\kern3pt#1\kern3pt}
\kern3pt\hrule}}\vrule}}

\def\ttilde#1{\raise2ex\hbox{${\scriptscriptstyle(}\!
\sim\scriptscriptstyle{)}$}\mkern-16.5mu #1}
\def\dddots#1{\raise1.5ex\hbox{$^{...}$}\mkern-16.5mu #1}
\def\siton#1#2{\raise1.5ex\hbox{$\scriptscriptstyle{#2}$}\mkern-16.5mu #1}
\def\upleftarrow#1{\raise1.5ex\hbox{$\leftarrow$}\mkern-16.5mu #1}
\def\uprightarrow#1{\raise1.5ex\hbox{$\rightarrow$}\mkern-16.5mu #1}
\def\upleftrightarrow#1{\raise1.5ex\hbox{$\leftrightarrow$}\mkern-16.5mu #1}
\def\bx#1#2{\vcenter{\hrule \hbox{\vrule height #2in \kern #1in\vrule}\hrule}}

\def\squiggle#1{\lower1.5ex\hbox{$\sim$}\mkern-14mu #1}

\def\narrower{\advance\leftskip by\parindent \advance\rightskip by\parindent}

\def\mbox#1#2{\vcenter{\hrule width#1in\hbox{\vrule height#2in
   \hskip#1in\vrule height#2in}\hrule width#1in}}
\def\eqsquare #1:#2:{\vcenter{\hrule width#1\hbox{\vrule height#2
   \hskip#1\vrule height#2}\hrule width#1}}
\def\inbox#1#2#3{\vcenter to #2in{\vfil\hbox to #1in{$$\hfil#3\hfil$$}\vfil}}
\def\strutdepth{\dp\strutbox}
\def\marbul{\strut\vadjust{\kern-\strutdepth\specialbul}}
\def\specialbul{\vtop to \strutdepth{
    \baselineskip\strutdepth\vss\llap{$\bullet$\qquad}\null}}
\def\Bcomma{\lower6pt\hbox{$,$}}    
\def\bcomma{\lower3pt\hbox{$,$}}    

\def\updots{\mathinner{\mskip 1mu\raise 1pt\hbox{.}
    \mskip 2mu\raise 4pt\hbox{.}\mskip 2mu
    \raise 7pt\vbox{\kern 7pt\hbox{.}}\mskip 1mu}}

\def\square{\kern1pt\vbox{\hrule height 1.2pt\hbox{\vrule width 1.2pt\hskip 3pt
   \vbox{\vskip 6pt}\hskip 3pt\vrule width 0.6pt}\hrule height 0.6pt}\kern1pt}

\def\lapp{\hbox{$ {     \lower.40ex\hbox{$<$}
                   \atop \raise.20ex\hbox{$\sim$}
                   }     $}  }
\def\rapp{\hbox{$ {     \lower.40ex\hbox{$>$}
                   \atop \raise.20ex\hbox{$\sim$}
                   }     $}  }
\def\Times{\times\hskip-2.3pt{\raise.25ex\hbox{{$\scriptscriptstyle^$}}}}

\def\rightonleft{\hbox{$ {     \lower.40ex\hbox{$\longrightarrow$}
                   \atop \raise.20ex\hbox{$\longleftarrow$}
                   }     $}  }
\def\svec#1{\skew{-2}\vec#1}

\topskip = 0.6 truein
\leftskip = 0.2 truein
\vsize=8.9truein
\hsize=6.3truein
\tolerance 10000
\hfuzz=20pt

\baselineskip 12pt plus 1pt minus 1pt
\centerline{CHIRAL SYMMETRY AND PARITY--VIOLATING MESON--NUCLEON
VERTICES\footnote{*}{Work
supported in part
by Deutsche
Forschungsgemeinschaft and
by Schweizerischer Nationalfonds.}
\footnote{$^{\#}$}{Invited talk presented
at the Workshop on ``Baryons as Skyrme Solitons'', Siegen, September,
1992.}}
\vskip 24pt
\centerline{Ulf-G. Mei{\ss}ner\footnote{$^\dagger$}{Heisenberg fellow}}
\vskip 12pt
\centerline{Universit\"at Bern}
\centerline{Inst.~f\"ur Theoret.~Physik}
\centerline{Sidlerstr.~5}
\centerline{CH-3012 Bern\ \ Switzerland}
\vskip 36pt
\centerline{ABSTRACT}
\medskip
{\noindent\narrower In this lecture, I review progress made in the
calculations of the parity-violating meson-nucleon interaction regions.  The
underlying framework is the topological chiral soliton model of the nucleon.
Emphasis is put on the computation of theoretically and experimentally
accessible nuclear parity violating observables.  I stress the importance of
the interplay of strong and weak interactions which makes this field
interesting and challenging. I also discuss recent developments pointing
towards the importance of strange quark admixtures in the proton
wave function.
\smallskip}
\goodbreak
\bigskip
\baselineskip 14pt plus 1pt minus 1pt
\noindent{INTRODUCTION}
\medskip
Our understanding of the hadronic weak interactions has progressed
considerably in the last two decades.  Still, the almost unique tool to study
the non-leptonic, strangeness conserving part of the weak Hamiltonian are
few-nucleon systems.  In general, nuclear parity-violating (pv) observables
cannot be calculated reliably enough so that we could deduce stringent limits
on the standard model from them.  Stated in another way: We are still far away
from extracting {\it e.g.\/} the Weinberg angle to some decent precision from
nuclear parity violation. On the other hand, there are now very sophisticated
parametrizations of the strong force between two nucleons available which
allow us to test our understanding of the hadronic weak interactions in terms
of meson exchanges.  Direct $W$- or $Z$-exchange between nucleons is wiped out
by the hard core of the $NN$-force, but there still remains a long-range
component \vfill \eject
\noindent{of} the weak interactions
between nucleons, which can be parametrized
in terms of pv meson-nucleon interaction vertices.  One way to calculate these
pv couplings is to make use of the quark model.$^1$  There is, however, a
considerable uncertainty in these calculations which stems from the fact that
the pertinent multiquark operators have to be calculated at low energies
$(E \ll M_{W,Z})$.
Gluonic corrections arise, and unavoidably one enters the
non-perturbative regime where the strong coupling constant $\alpha_s(q^2)$
becomes larger than unity.  These problems are most pronounced in the case of
the pion which dominates the long-range part of the pv potential.  The
Goldstone-boson-like character has always posed problems for quark-model
practitioners.  Quite contrary, the recently popular topological soliton
models of the nucleon like the Skyrme model$^2$ and generalizations
thereof$^3$ naturally incorporate the pseudoscalar as well as the low-lying
vector multiplets.  Here we have reached our starting point for a calculation
of the parity-violating meson-nucleon couplings and form
factors.$^{4,\,5,\,6,\,7}$  The soliton approach to the nucleon is far from
being perfect, but it has the conceptual advantage that it allows for a
simultaneous calculation of the strong and weak interaction regions, a point
which is generally overlooked by quark model enthusiasts.  Furthermore, nuclear
parity violation can also be used as a testing ground to find out the
limitations of the
soliton scenario --- often more can be learned from the failures of a model
than from its successes.

Another interesting aspect of nuclear parity violation is the quest for
finding few-nucleon systems which can be calculated with some reliability and
where the experimenters have a change of detecting a clear signal.  Here, I
will focus on two rather different systems.  In proton-proton scattering, one
can observe longitudinal asymmetries of the order $10^{-7}$, which appear to
be awfully small.  However, progress in experimental techniques now allows for
experiments with an accuracy of $\delta A_L\simeq \pm 1.0\cdot 10^{-8}$ and
therefore a fairly sensitive test of the meson-exchange picture underlying the
theoretical description of this process.  A very different system is the
nucleus ${}^{18}$F, in which nuclear amplification takes place and the
observed circular polarization of emitted $\gamma$-rays is of the order
$2\cdot 10^{-4}$.  Luckily for the theorists, the $\beta$-decay of the
daughter nucleus ${}^{18}$Ne allows one to gauge the rather involved shell
model calculations,$^8$ although sceptical minds tend to look at these
calculations with a certain dose of disbelieve.  As we will see, too few
``good'' nuclear systems are considered at present and therefore the
restrictions on the pv meson-nucleon couplings are by far too soft.

Finally, during the last year, effective field theory methods have
been used to gain further insight  into the strength of the pv meson--nucleon
couplings.$^9$ These results seem to indicate a large enhancement
from operators involving strange quarks to various coupling constants.
Furthermore, some couplings not considered so far (like {\it e.g.}
the pv $\pi N \gamma$ vertex) might be of importance. I will discuss
these topics in the end of this lecture.
\goodbreak
\bigskip
\noindent{PV MESON-NUCLEON INTERACTION REGIONS}
\medskip
In the meson-exchange parametrization of the weak nuclear force, one usually
only considers the exchange of charged pions and the vector mesons $\rho$ and
$\omega$.  CP invariance does not allow for the coupling of neutral scalar or
pseudoscalar mesons to nucleons, eliminating the infamous scalar mesons, the
$\eta$, $\eta'$ and the $\pi^0$ (the $\delta^\pm$ are considered a form factor
corrections to the $\pi^\pm$-exchange).  Then there remains the $\phi(1020)$
--- its coupling to the nucleon is generally supposed to be OZI-suppressed and
not considered.$^{10}$  This might, however, be a too simplistic approach in
light of the discussion surrounding the admixture of strange operators into
the proton's wavefunction. At present no final conclusion can be drawn
and I will make life easy on us and neglect the $\phi$ for the time being.
I will pick up this theme in the final section.

Unavoidably I will have to define the basic couplings which parametrize the pv
nuclear potential.  For the pion, there is only a $\Delta I = 1$ (isovector)
coupling (to first order in the pion field)
$${\cal L}^{\rm pv}_{\pi N} = - {G_\pi (q^2)\over \sqrt{2}}{E\over M_N}
\chi^\dagger_f\left( {\svec\tau} \times {\svec\pi}\right)_3 \chi_i\eqno(1)$$
with $\chi_{i,f}$ denoting nucleon spinors, $q^2$ the invariant momentum
transfer squared
at the $\pi N$-vertex and I have (for simplicity) given the
non-relativistic reduction of this vertex.  In the case of the $\omega$,
$\Delta I = 0$ and $\Delta I = 1$ couplings are possible,
$${\cal L}^{\rm pv}_{\omega N} = \chi^\dagger_f \left[ h^0_\omega (q^2) +
h^1_\omega(q^2) \tau^3\right]\left[ {E\over M_N} {\svec\sigma}_T
+{\svec\sigma}_L\right]\cdot {\svec\omega}\chi_i \eqno(2)$$
with $E= \left( M^2_N + {\svec q}^2/4\right)^{1/2}$ and ${\svec\sigma}_{L,T}$
the longitudinal and transverse spin-operator, respectively.  For the $\rho$,
one has isoscalar, isovector and isotensor vertices
$$\eqalign{
{\cal L}^{\rm pv}_{\rho N} &= \chi^\dagger_f \left[ h^0_\rho (q^2) \tau^a +
h^1_\rho (q^2) \delta^{a3} + {h^2_\rho (q^2) \over 2\sqrt{6}} \left( 3\tau^3
\delta^{a3} - \tau^a\right)\right] \cr
&\times \left[ {E\over M_N} {\svec\sigma}_T + {\svec\sigma}_L \right]\cdot
{\svec\rho}^a \chi_i - {iE\over 2 M^2_N} h^{'1}_\rho (q^2)
\chi^\dagger_f{\svec\sigma} \cdot{\svec q}\left(
{\svec\tau}\times{\svec{\rho^0}}\right)_3 \chi_i  \ \ .\cr} \eqno(3)$$
Generally, the coupling $h^{'1}_\rho$ is neglected,$^{11}$ but I will not
follow this historical path here.  From \hbox{Eqs.} (1) -- (3) it becomes
obvious that the pv interaction regions are characterized by coupling
constants $h_M = h_M(q^2=0)$ and form factors $F^{\rm pv}_{M} (q^2) = h_M
(q^2)/h_M$ (in case of the pion, I use $G_\pi \equiv h_\pi$).

In the topological chiral soliton model$^{12}$ underlying the calculation of
the pv vertices, nucleons arise as solitons of a non-linear meson theory.
This non-linear meson theory is constructed in harmony with chiral symmetry
and anomaly constraints and all its parameters are fixed from mesonic
reactions like {\it e.g.\/} $(\rho^0\to\pi^+\pi^-$, $\omega\to
\pi^+\pi^-\pi^0$, $\omega\to \pi^0\gamma,\ldots$~~.  The Lagrangian and its
parameters are completely determined in the meson sector, and the calculation
of nucleon properties proceeds without any new parameters, {\it i.e.\/} no
fudging is possible!  That is certainly an appealing aspect of the soliton
approach to the nucleon and it poses several restrictions.  Of course, the
model does not perfectly predict all nucleon properties.

Now: How can we calculate the pv couplings appearing in \hbox{Eqs.} (1) --
(3)?  For that, we consider the current $\times$ current form of the weak
Hamiltonian with the currents being of ($V$--$A$)-type.
To pick out the pv pieces,
consider the $\Delta I = 0,1$ or 2 components of products like $V_\mu A^\mu$
and $I_\mu A^\mu$,
with $V_\mu$ the vector, $I_\mu$ the isoscalar and $A_\mu$ the axial current.
These currents are
already given in terms of the meson fields which make up the soliton, and
their explicit expressions can be found {\it e.g.\/} in \hbox{Ref.} [5].  One
then makes use of the ``background-scattering'' method, which amounts to an
expansion of the meson fields around the soliton background.  For the pion, we
write$^{13}$
$${\svec\pi} = {\svec \pi}_S + {\svec\pi}_f \eqno(4)$$
and similarly for $\rho$ and $\omega$.  ${\svec\pi}_S$ is the ``hard''
component of the pion field making up the soliton and ${\svec\pi}_f$ a small
pionic fluctuation (``soft'' component).  Inserting the expressions (4) into
the soliton currents and these into the weak Hamiltonian, all one has to do is
to find the terms linear in ${\svec\pi}_f$ (or ${\svec\rho}_f$ or $\omega_f$).
Quantizing the respective operators which are given in terms of the collective
variables $(A,\dot A)$,$^{2,\,3}$ one can immediately read off the coupling
constants and form factors for the meson under consideration.  In particular,
one cannot only construct pv meson-nucleon vertices, but also the equivalent
pv $N\Delta$-transition couplings.  I will come back to this point later on.
For details, the interested reader should consult \hbox{Refs.} [5,6].  I will
not give any explicit formula here, but rather make a few comments on the
results of \hbox{Ref.} [5].  First, the pv $\pi N$ coupling is completely
dominated by the neutral current contribution.
\bigskip
$$\hbox{\vbox{\offinterlineskip
\def\strut{\hbox{\vrule height 15pt depth 12pt width 0pt}}
\hrule
\halign{
\strut\vrule# \tabskip 0.1in &
\hfil#\hfil &
\vrule# &
\hfil#\hfil &
\hfil#\hfil &
\hfil#\hfil &
\hfil#\hfil &
\hfil#\hfil &
\vrule# \tabskip 0.0in
\cr
&  && KM & DDH & DZ & AH & RR & \cr
\noalign{\hrule}
& $\tilde F_\pi$ && 0.6 & 10.8 & 3.1 & 5.0 & 0.0$\to$27.1 &
\cr
& $F_0$ &&   5.9 &   15.9 &  11.5 & 8.0 &--15.9$\to$43.0 &
\cr
& $F_1$ &&   0.2 &   0.3 & --0.5 & 0.3 & --0.1$\to$0.6 &
\cr
& $F_2$ &&   5.3 &  13.3 &   9.3 & 9.8 &   10.6$\to$ 15.3 &
\cr
& $H_1$ &&   1.7 &  0.5 & 0.0 & 0.0 & --- &
\cr
& $G_0$ &&  24.5 &   8.0 &  16.3 & 27.0& --23.9$\to$ 43.1 &
\cr
& $G_1$ &&   4.1 &   4.8 &   9.2 & 10.0 & --3.3$\to$ 8.0 &
\cr\noalign{\hrule}
}}}$$
\smallskip
{\noindent\narrower Table: Effective weak meson-nucleon coupling constants
in units of $10^{-7}$. We present the result of the soliton model
calculation of Kaiser and Mei{\ss}ner(KM) [5,6] together with the quark
model results of Desplanques {\it et al.}(DDH) [1] as well as Dubovik and
Zenkin(DZ) [16]. The value for $h^{'1}_\rho$ in the column DDH is taken from
Holstein's calculation in ref.[11]. The ``reasonable ranges''(RR) defined by
DDH are also given. The column AH gives the best fit values of Adelberger
and Haxton.$^{17}$\smallskip}
\goodbreak
\medskip
\noindent The charged current
contribution can be estimated in the factorization approximation,
$G^{CC                                                                  }_\pi
= \cos^2\theta_c <\pi| A_\mu |0> <p |
V_\mu| n> = G_F \cos^2 \theta_cf_\pi (M_n - M_p)$.  The electromagnetic
mass difference of the neutron and the proton is well-reproduced in the
model,$^4$ whereas a strong part of $M_n - M_p$ is somewhat
underestimated.$^{14}$.  Taking as an upper limit the empirical value $M_n -
M_p \simeq 1.3$~MeV, we find $A=G^{NC}_{\pi}\big/ G^{CC} \sim 13.5$,
consistent with previous estimates$^{15}$ and the quark model calculations of
\hbox{Ref.} [1] $(A\simeq 24)$.  The numerical value for the effective
pion-exchange coupling, $\tilde F_\pi = g_{\pi NN} G_\pi\big/ \sqrt{32}$ with
$g_{\pi NN}$ the strong $\pi N$ coupling, is considerably smaller in the
soliton model than in the quark model, we find $\tilde F^{\rm sol}_\pi = 0.6$
ve
rsus
$\tilde F^q_\pi = 10.8$,$^8$ or $\tilde F^q_\pi = 3.1$\ $^{16}$ (in units of
$10
^{-7}$).

For the vector meson couplings, the results are less different.  Using the
standard definitions $F_i = - g_{\rho NN} h^i_\rho/2$, and $G_i = - g_{\omega
NN} h^i_\omega/2$ and $H_1 = - g_{\rho NN} h^{'1}_\rho/4$, we find that the
soliton and the quark model predict the following pattern for the
$\rho$-couplings:  $F_0\rapp F_2  \gg F_1$.  The absolute values of the
constant
s
$F_i$, are, however, reduced in the soliton approach.  For $H_1$, the soliton
model predicts a value three times as large as the quark model.$^{11}$  In the
case of the $\omega$, all calculations give $G_0>G_1$, but $G_0$ is
considerably enhanced in the soliton approach and close to the ``best-fit''
estimate of Adelberger and Haxton.$^{17}$ These results are summarized
in the table.

As already stated, the calculation of pv $N\Delta M$ \ $(M = \pi,\rho,\omega)$
transition vertices proceeds along the same lines and only differs at the step
when one quantizes the collective coordinates.  In our model, the pv $\pi
N\Delta$
coupling has $\Delta I = 0$ and 1 components and reads non-relativistically
$${\cal L}^{\rm pv}_{\Delta N\pi} = \left( {E\over 4 M^3_N}\right) \left[
\delta^{ab} h^0_\Delta(q^2) +\epsilon^{3ab} h^1_\Delta (q^2)
\right]\chi^\dagger_\Delta {\svec S} \cdot {\svec q} {\svec\sigma}\cdot {\svec
q} T^a \chi_N \pi^b + \hbox{h.c.} \eqno(5)$$
with ${\svec S}$ and ${\svec T}$ the conventional $N\Delta$ transition spin
and isospin operators, respectively.  It is easy to convince oneself that
$h^0_\Delta(q^2) = 0$ in this model,$^6$ naively, a non-vanishing isoscalar
$\pi\Delta N$-vertex would lead to a non-zero CP-violating $\pi N$-vertex (a
more fool-proof argument is given in \hbox{Ref.} [6]).  The isovector vertex
does not vanish, and for the ``minimal'' model$^{18}$ we find
$$h^1_\Delta(0) \big/ G_\pi = 1.10\ \ .\eqno(6)$$
The presence of $\pi\rho\omega$-correlations in the effective action tends,
however to decrease this ratio.  For the $V\Delta N$-couplings, we find
$(V=\rho$ or $\omega)$:
$$h^i_{\rho \Delta N} (q^2) = {3\over\sqrt{2}} h^i_\rho(q^2)\qquad
(i = 0,'1,2) \ , \qquad
h^1_{\omega\Delta N} (q^2) = {3\over \sqrt{2}} h^1_\omega (q^2) \eqno(7)$$
and $h^1_{\rho\Delta N}(q^2) = h^0_{\omega\Delta N}(q^2) \equiv 0$.  These
predictions are insofar interesting since in the seventies it was argued that
{\it e.g.\/} the $\rho\Delta N$-couplings are negligible$^{19}$ --- quite in
contrast to our results. A recently performed quark model calculation
by Feldman {\it et al.}$^{29}$ along the lines of DDH gives results rather
different from the soliton model predictions. The source of these discrepancies
is not yet understood.
In a similar fashion, one easily derive the corresponding $M \Delta \Delta$
vertices,
$$ \eqalign{
& G_{\pi \Delta \Delta} (q^2) = G_\pi (q^2) \cr
& h^i_{\rho \Delta \Delta} (q^2) = {1 \over 5} h^i_\rho (q^2) \, \,
(i =0, '1, 2) \, \, ; \, \,
 h^1_{\rho \Delta \Delta} (q^2) =  h^1_\rho (q^2) \cr
& h^0_{\omega \Delta \Delta} (q^2) = h^0_\omega (q^2) \, \, ; \, \,
 h^1_{\omega \Delta \Delta} (q^2) = {1 \over 5} h^1_\omega (q^2) \cr}
\eqno(8) $$

The calculation of the associated weak form factors proceeds in a
straightforward way.$^6$  In Fig.~1 we show the weak $\pi N$ form factor
$G_\pi (q^2)$ in comparison with the equivalent strong form factor
$G_{\pi NN}(q^2)$
as well as the monopole with cut-off $\Lambda =1$~GeV.  As it turns out, all
form factors can be fitted by monopoles at low ${\svec q}^2$,
$$F^M ({\svec q}^2) = h_M {\Lambda^2\over\Lambda^2 + {\svec q}^2} \eqno(9)$$
\midinsert
\vskip 10.5truecm
{\noindent\narrower \it Fig.~1:\quad
The weak $\pi N$ form factor $G_{\pi} (q^2)$ in comparison to its
strong counterpart $G_{\pi N N} (q^2)$ and a monopole fit with a
cut--off $\Lambda = 1$ GeV (solid line).
\smallskip}
\vskip -0.5truecm
\endinsert
\noindent
and are very similar to the respective strong form factors.  This is the first
time that   such a calculation has been performed and its result can be
understood
as follows:  The intrinsic scale of the meson-nucleon interaction regions is
set by the topological baryon charge radius, $r_B\simeq 0.5$~fm.  From that,
one can deduce a cut-off scale $\Lambda \simeq \sqrt{6}\,/r_B \simeq 1$~GeV.
It is, however, not that simple because the dynamical treatment of the vector
mesons modifies this result.  Defining by $R_M$ the ratio of the (averaged)
weak to strong $MN$ cut-offs (all form factors of monopole type), we find
$$R_\pi = 1.15\ \ ,\quad R_\rho = 0.91\ \ ,\quad R_\omega = 0.77 \eqno(10)$$
which justifies within the accuracy of the model the assumption of taking the
same form factors for the strong and weak vertices as it was done {\it e.g.\/}
by Driscoll and Miller$^{20}$ in their study of the pv $pp$-interaction.

Another topic which can be discussed in the framework of the chiral
soliton model are the corrections of pv two--pion exchange. One
motivation to do this is that correlated $2 \pi$--exchange gives rise
to the intermediate range attraction of the parity--conserving $NN$
interaction. Furthermore, recent investigations point towards the
importance of pv $2 \pi$ exchange even below production threshold.$^{31}$
Using the soliton model, Norbert Kaiser and I have shown that the
inclusion of pion loops gives the intermediate range attraction
with just the right strength as compared to the Paris potential.$^{32}$
Similarly, we have worked out corrections to the various pv $\rho N$
couplings.$^{33}$ The effects of irreducible two--pion corrections
are generally small, of the order of $10 \ldots 20$\%. This is
in agreement with older dispersion--theoretical investigations.$^{34}$
So we finally have all the tools at hand to make contact to experiment.
\goodbreak
\bigskip
\noindent{PARITY-VIOLATION IN PROTON-PROTON SCATTERING}
\medskip
\nobreak
The simplest system in which one can probe certain components of the weak pv
inter-nucleon force is the two nucleon system.  By scattering polarized
protons off a hydrogen target, parity violation shows itself in a non-vanishing
longitudinal asymmetry,
$$A_L = {\sigma^+ - \sigma^-\over\sigma^+ + \sigma^-} \eqno(11)$$
assuming a 100\% longitudinal polarization of the beam and having taken care
of the Coulomb-corrections $\sigma^\pm$ are the cross-sections for scattering
positive/negative helicity protons from an unpolarized target.  The calculation
of this process in the DWBA as pioneered by Brown {\it et al.\/}$^{21}$ goes
as follows.  One splits the total scattering amplitude ${\cal F}_{ss'}$ into a
strong and weak part
$${\cal F}_{ss'} = F_{ss'} + f_{ss'} \eqno(12)$$
for total spins $s$ and $s'$.  Now it is of utmost importance to take into
account the strong distortions, {\it i.e.\/} calculating the weak scattering
amplitude with distorted waves $\psi^{(-)}_s$ and $\psi^{(+)}_s$, {\it i.e.\/}
$f_{ss'} = <\psi^{(-)}_s|V_{\rm pv}| \psi^{(+)}_s>$                     with
$V_{\rm pv}$ the pv one-meson-exchange potential.  It should be pointed out
that the strong distortions govern the energy-dependence of the analyzing
powers $A_L$.  Recently, Driscoll and Miller$^{20}$ have done the most complete
calculation based on the Bonn-Potential$^{21}$ for the strong force and an
equivalently constructed weak potential with the pv couplings taken from the
quark model$^1$ and using the same vertex functions for the weak and strong
form factors.  I should point out here that for obvious reasons there is no
pion contribution to this process and one essentially tests the vector-meson
couplings $h^{pp}_\rho = h^0_\rho + h^1_\rho + h^2_\rho\big/\sqrt{6}$ and
$h^{pp}_\omega = h^0_\omega + h^1_\omega$.

Recently, Doug Driscoll and I have repeated this calculation$^{22}$ by
including the soliton model predictions $h^{pp}_\rho = -5.15\cdot 10^{-7}$ and
$h^{pp}_\omega =-8.20\cdot 10^{-7}$.  The resulting curve for $A_L$ is shown
in Fig.~2, for the quark$^{1,16}$
and the soliton model.$^{22}$  The shape of the
curve as predicted by the soliton model follows closer the empirical trend
suggested by the low-energy data.$^{23}$ In fact, a $\chi^2$ calculation
for the three curves shown in fig.2 gives $\chi^2 = 34/3$ (DDH),
$\chi^2 = 26/3$ (DZ) and $\chi^2 = 8/3$ (KM) as disussed in ref.26 (at
that time, the Bonn result was not available).
Also, the maximum at $p_{\rm
lab}=0.95$~GeV/c is flatter than in the calculation using the quark model
parameter.  Furthermore, the energy at which the asymmetry changes sign is
larger than the quark model predicts, which can be traced back to the fact
that in the soliton model $h^{pp}_\omega>h^{pp}_\rho$, in contrast to the
quark model with $h^{pp}_\omega <h^{pp}_\rho$.  Of particular interest is the
value of $A_L$ at 222~MeV.  This is the energy selected for an upcoming $pp$
parity violation measurement at TRIUMF because $\delta({}^1S_0) +
\delta({}^3P_0)=0$ at this energy and the $j=0$ contribution to the analyzing
power consequently vanishes. The measurement of the dominant $j=2$
contribution gives a different combination of $h^{pp}_\rho$ and
$h^{pp}_\omega$ than the $j=0$ contribution to $A_L$, which is already
measured at 15 and
45~MeV.$^{23}$.  The predictions using the quark$^1$ and soliton
model$^{5}$ weak parameters, respectively, differ by $\Delta A_L = 4.6\cdot
10^{-8}$. To be more precise, the various predictions are:
$$ A_L ({\rm DDH}) = 5.0 \cdot 10^{-8}\, \, , \, \,
   A_L ({\rm DZ}) = 2.6 \cdot 10^{-8}\, \, , \, \,
   A_L ({\rm KM}) = 3.7 \cdot 10^{-9}   \eqno(13)   $$
The projected long-term accuracy of the upcoming TRIUMF experiment
is $\left( \delta A_L\right)_{\rm stat} \simeq \pm 1\cdot 10^{-8}$,$^{24}$
which should be sufficient to discriminate between these two predictions.
Notice that a similar experiment is also planned at COSY.$^{30}$
This experiment should set rather stringent limits on some combinations of the
pv $\rho N$ and $\omega M$ couplings. To stress it again, the $pp$ system
is a particularly good example of the interplay of weak and strong
interactions and it is therefore mandatory to treat both of them consistently
(for further discussion, see \hbox{Refs.} [17,20]).
A possible loophole to all of this will be discussed in the last section.
\goodbreak
\bigskip
\midinsert
\vskip 13.0truecm
{\noindent\narrower \it Fig.~2:\quad
Parity-violating asymmetry in $pp$-scattering. The solid
line gives the prediction based on the weak couplings as given by the
soliton model,$^{22}$
whereas the dashed and dashed-dotted lines are based
on the quark model calculations of refs.1 and 16, respectively.
\smallskip}
\vskip -0.5truecm
\endinsert
\noindent PARITY VIOLATION IN ${}^{18}$F AND DEUTERON PHOTODISINTEGRATION
\medskip
\nobreak
The nucleus ${}^{18}$F is what I called a ``good system'' before.  It
exhibits ``nuclear amplification'' in that it has two close-by levels of
opposite parity which are separated by only 39~keV (the next level which could
mix with these is approximately 2~MeV away) and the dominant E1-transition
from the level at 1.081~MeV to the ground state is suppressed, which leads to
$|M1/E2| \simeq 112$.  The M1-transition is, of course, only possible because
of the mixing of the opposite parity-levels.  Altogether, this amounts to an
amplification of approximately $(2/0.039)*112\simeq 6\cdot 10^3$ (for further
details, see \hbox{Ref.} [17]).  Theoretically, one can calibrate the
shell-model calculation to extract the pv circular polarization from the
$\beta$-decay of ${}^{18}$Ne, because the pion-exchange of this $\beta$-decay
up to an overall isospin rotation,$^{25}$ and therefore calculation and
measurement of ${}^{18}$Ne\ $(0^+1)\to {}^{18}$F\ $(0^-0)$ \ $\beta$-decay
serves as a ``gauge'' for the accuracy and amounts effectively to a large
model-independent limit on the weak pion decay constant.  The latter dominates
completely this $\Delta I=1$ pv observable, and one can deduce a limit on
$\tilde F_{\pi}, $\ $\tilde F_\pi\le 3.4\cdot 10^{-7}$. Here, we have used the
experimental
circular polarization, $|P_\gamma({}^{18}\hbox{F})| = (0.17 \pm 0.58)\cdot
10^{-3}$.  The quark model prediction of \hbox{Ref.} [1], $\tilde F_\pi = 10.8
\cdot
10^{-7}$, is clearly in contradiction to this result.

What does the soliton model give?  Of course, $\tilde F_\pi$ is considerably
reduced,
so we expect a smaller asymmetry.  The vector meson contribution is enhanced,
and taking nuclear structure calculation from \hbox{Ref.} [17], we predict
$P_\gamma({}^{18}\hbox{F}) = 2.2\cdot 10^{-4}$, not far from the central value
of the experiment.  We should, however, not put too much emphasis on this
closeness of the experimental and theoretical number, but rather state that
the strength of the pv $\pi N$ coupling should still be considered as the main
theoretical puzzle.  I am sure that $\tilde F_\pi$ should come out smaller than
in
\hbox{Ref.} [1], but whether it is as small as predicted by the soliton model
can only be checked if more theoretical and experimental information on the
$\Delta I = 1$ part of the pv nuclear force are available.  One particularly
interesting candidate to study in more detail would be the reaction ${\svec
n}+p\to d+\gamma$ or the inverse process $ \vec \gamma + d \to n + p$.
A calculation of the circular asymmetry as a function of the photon
energy has been performed some time ago by Oka.$^{27}$ I have used this
calculation in Ref.28  to investigate the sensitivity of the circular
asymmetry $A_L$
to the various pv couplings. Considering photon energies
below 30 MeV, $A_L (\omega)$ increases linearly with energy
when one uses the quark model couplings of DDH or DZ, with the slope
determined by the strength of the pv $\pi N$ coupling.
For the DDH-case
the pion contribution is completely dominant for all energies, whereas
for the DZ-parameters the reduced $\pi NN$ strength leads to an overall
decrease of $A_L(\omega)$. For the soliton model, however, things are
significantly different. First, between $1$ and $20\,MeV$, $A_L(\omega)$
shows a flat minimum at about $\omega_L \approx 12\,MeV$ and only after
$\omega_L \geq 20\,MeV$ a gradual rise in $A_L(\omega)$ sets in. Also, the
overall magnitude of the effect is an order of magnitude smaller for the
weak parameters predicted by the soliton model. It would be worthwile to
measure the asymmetry say at $10$ and $20\,MeV$ incident energy,
although the effect is small, the tremendously different slope of
$A_L(\omega)$ should be detectable in an dedicated experiment. Of course,
as already mentioned, a more thorough theoretical study has also to be
done. First, a more consistent calculation employing $e.g.$ the
Bonn-potential and the equivalently constructed weak potential should be
performed. Second, the effects of meson-exchange curents, which play an
important role in the accurate description of the deuteron properties
have to be included. Therefore, these results
should only be considered as a guide, but the trends exhibited will
certainly not be wiped out by a more elaborate calculation.
A more detailed discussion is given in Ref.28.

In the last section, I will discuss some medium renormalization effects
which might come to the rescue of the large value for the pv pion--nucleon
coupling as predicted by DDH. However, for the deuteron photodisintegration
process just discussed, such a renormalization cannot be operative
since the deuteron is essentially an ensemble of two free nucleons.
\goodbreak
\bigskip
\noindent{THE NUCLEON ANAPOLE MOMENT}
\medskip
Apart from the electric dipole moment, there is one other pv coupling of the
photon to nucleons (spin-1/2 fields), the so--called anapole moment. It has
recently attracted new interest$^{35}$  since its contribution might be
enhanced
considerably in nuclei, similar to the case of $^{18}$F just discussed. For
on--shell nucleons, current conservation and Lorentz invariance require that pv
corrections to matrix elements of the electromagnetic current take the form
$$ < N(p') | J^{\rm em}_{\mu , pv} (0) | N(p) > = { a(q^2) \over M_N^2} \bar u
(p') [ \gamma_\nu q^\nu q_\mu - q^2 \gamma_\mu ] \gamma^5 u(p)   \eqno(14)  $$
with $q^2 =(p' - p)^2$. In the Breit frame, where the photon transfers no
energy, this matrix element reads
$$ < N(\vec{q}/2) | J^{\rm em}_{\mu , pv} | N (- \vec{q}/2) > = { E \vec{q}^{\,
2} \over
M_N^3} a(\vec{q}^{\, 2} ) \chi_f^\dagger \vec{\sigma}_T \chi_i    \eqno(15)
$$
with $\vec{\sigma}_T  =\vec{\sigma} - \hat{q} \vec{\sigma} \cdot \hat{q}$ the
transverse spin operator and $a (q^2 )$ the nucleon anapole form factor.
The anapole moment has isoscalar and isovector components,
$$ a (0) = a_S (0) + a_V (0)  \tau_3           \eqno(16)    $$
In the soliton model, one can easily calculate the anapol moment and form
factor.$^{36}$ For that, one identifies the matrix  element in (15) with the
Fourier transform of the pv electromagnetic current. For the usual hedgehog
ans{\"a}tze, its has the general form
$$ \vec{J}_{pv} ( \vec r ) = \Gamma_1 (r) \, \vec{\sigma} + \Gamma_2 (r)
\, \hat{r} \vec{\sigma} \cdot \vec r                \eqno(17)    $$
with $\Gamma_{1,2} (r)$ functions of the various meson profiles whose
explicit form we do not need here. However, one immediately encounters a
difficulty. Current conservation demands $\Gamma_1 ' (r) + \Gamma_2 ' (r)
+ 2 \Gamma_2 (r) / r = 0$, where the prime denotes differentiation with
respect to $r$. This condition is not met. Interestingly, if one switches off
the $\rho$--meson fields and considers the so--called $\omega$--stabilized
Skyrmion$^{37}$, then the divergence condition is fulfilled. This peculiar
behaviour might be traced back to the fact that in the isoscalar channel one
has exact vector meson dominance (VMD) but not in the isovector one (compare
the discussion of Hwang and Nigoyi$^{38}$ of VMD and gauge invariance for pv
photon--nucleon couplings). To get an idea of the size of the anapole moment,
let me crudely rstore gauge invariance by subtracting the pieces which
violate current conservation. In that case, the "minimal" model gives $a_s
(0) = 4 \cdot 10^{-8} $, and the  extension of the pv $\gamma N$ vertex is
given by a mean square radius of about 0.4 fm corresponding to  a monopole
form factor with a cut--off of $\Lambda = 1.23$ GeV. At present, I can not
offer a solution to the problem concerning the violation of current
conservation, but I suspect that is it related to the rather crude
quantisation procedure used (which is known to do harm to e.g. the chiral
algebra of the charges$^{38}$).
\bigskip
\noindent{RECENT DEVELOPMENTS}
\medskip
There are some recent developments (partly outside the soliton model) which
indicate some interesting new effects and might lead to a reconsideration of
some topics discussed so far. The first one is due to a calculation of Dai,
Savage, Liu and Springer.$^9$ They calculate an effective Hamiltonian for
$\Delta I = 1$ nuclear parity violation, including the effects of the heavy
quarks $s,c$ and $b$. At the scale of the W-boson mass, the pv $\Delta I = 1$
Hamiltonian is, of course, well known and given in terms of eight four--quark
operators with known Wilson coefficients. Integrating out the $b$ and the $c$
quark successively, one has a tower of effective theories. For each of these,
the anomalous dimension matrix is calculated to one loop in the QCD
corrections and the effective field theories are matched. By this procedure,
one can finally go down to the hadronic scale of $\Lambda_\chi = 1$ GeV and
compare the Wilson coefficients $C_i (\Lambda_\chi )$ with the original ones,
$C_i (M_W)$. The important observation made in ref.9 is that the operators
involving strange quarks are substantially larger than the ones involving
only the up and down quarks, approximately
$$ {C_i^{\rm strange} ( \Lambda ) \over C_i^{\rm non-strange} ( \Lambda ) }
\sim {1 \over \sin^2 \Theta_W}  \sim 5                    \eqno(18) $$
The authors of ref.9 did not compute hadronic matrix elements at the scale
$\Lambda_\chi$, but resorted to the meson--exchange picture and large $N_c$
arguments. In that case ($N_c \to \infty$), factorization can be justified
and one finds for the $\rho^0 N$ pv matrix element
$$\eqalign{
< \rho^0 N | H_{pv}^{\Delta I = 1}  | N >  = {1 \over 3} G_F \sin^2
\Theta_W f_\rho {\epsilon_\rho^*}^\mu \lbrace &-0.95 <N | \bar{u} \gamma_\mu
\gamma_5 u + \bar{d} \gamma_\mu \gamma_5 d | N > \cr &+ 13.4
< N | \bar{s}  \gamma_\mu \gamma_5 s | N > \rbrace \cr}
\eqno(19) $$
Consider first the case were the strange matrix element vanishes (like in
DDH). In that case, one finds $h_\rho^{(1)} = -1.9 \cdot 10^{-8}$, quite
consistent with the DDH value. However, the large relative factor in front of
the new, un--colored strange contribution can easily alter this result by an
order of magnitude.  Combining the EMC--data and hyperon decay rates, one has
$ <N | \bar{u} \gamma_\mu
\gamma_5 u + \bar{d} \gamma_\mu \gamma_5 d | N > \simeq -
< N | \bar{s} \gamma_\mu \gamma_5 s | N >
\simeq -(0.2 \pm
0.1) S_\mu$ with $S_\mu$ the nucleon spin vector. In this case,
$h_\rho^{(1)} = -2.9 \cdot 10^{-7}$, which is an enhancement of a factor 15.
If that were true, all previous estimates of pv meson--nucleon couplings can
be off the mark by large factors. However, we should not forget that in last
years many of the matrix elements which indicated a large contribution of the
stramge quark sea to various nucleon properties have been tamed, the prime
example being the famous $\pi N$ $\Sigma$ term.

In a similar fashion, Kaplan and Savage$^{40}$ have recently reanalyzed  the
pv pion--nucleon couplings making use of baryon chiral perturbation theory.
They have derived the most general pv and CP--conserving effective
pion--nucleon--photon Lagrangian to first order in derivatives and first
order in the photon field and to all orders in the pion field. This effective
Lagrangian is parametrized by a few coupling constants, which are labelled
$h_V^{0,1,2} \, , \, h_A^{1,2}$ and $h^1_{\pi NN} = G_\pi$. Apart from the
standard pv pion--nucleon coupling (discussed before), the authors of ref.40
mainly concentrate on the novel pv $\gamma \pi NN$ and the pv $\pi \pi NN$
vertices (the latter one has been already been considered by nuclear
theorists in the seventies). Three different methods are used in ref.40 to
estimate the strength of these coupling constants, namely factorization,
dimensional analysis and relations to $\Delta S = 1$ hyperon decay matrix
elements. From these methods, the dimensional analysis is considered most
reliable. The most interesting results of this are 1) a large contribution of
the strange quarks to $G_\pi$ (together with a large value for this
coupling),
2) a sizeable strangeness enhancement for the pv $\pi \pi NN$ coupling
$h_A^1$ and 3) a large value for the strength of the pv $\gamma \pi NN$
coupling. Taking these estimates face value, drastic consequences would
arise. First, in the case of the $^{18}$F experiment, interference between
the one--pion exchange (considered so far) and  the novel $\gamma \pi NN$
vertex might complicate the analysis of the data and ultimately relax the
bound on $G_\pi$. Similarly, for the planned TRIUMF and COSY experiments
measuring parity violation in $pp$ scattering at 230 MeV, one would have to
consider  two--pion exchange, not only the conventional one arising from
e.g. intermediate $\Delta$ resonances, but also the one due to the large pv
$\pi \pi NN$ coupling. However, before jumping too far, one should not forget
that the results of ref.40 should be considered indicative -- more elaborate
calculations of the hadronic matrix elements are necessary (using e.g. lattice
methods) and also more complex nuclear structure calculations involving these
novel couplings have to be performed before one can draw a final conclusion.
For more details on these topics, please consult ref.40.
\bigskip
\noindent{MEDIUM RENORMALIZATION OF $G_\pi$ ?}
\medskip
\goodbreak
There exist ample evidence that suggests scale changes of fundamenral
properties of nucleons in nuclei. Some pertinent examples are the first EMC
effect, the quenching of the axial--vector coupling constant $g_A$ in nuclear
$\beta$--decay or the behaviour of the longitudinal and transverse strength
functions in quasi--elastic electron scattering off nuclei. These effects are
there and they are important, but thier origin still remains to be explained
in a consistent treatment of many--body effects and fundamental scale changes
of the nucleon properties. The chiral soliton  model allows to systematically
investigate the constraints from chiral symmetry on such possible medium
modifications.$^{41,42}$ The basic idea is the following: In the soliton
model, baryon properties are fixed once the mesonic input is determined. We
know, however, that meson masses and coupling constants change in the
baryon--rich environment.$^{43}$ This immediately leads to density or
temperature--dependent nucleon properties.$^{44}$ For the meson sector, I
will use here results from the Nambu--Jona-Lasinio model which have been
obtained in collaboration with V{\'e}ronique Bernard.$^{42}$ For not too large
densities $\rho$, one finds for the pion decay constant and the vector and
scalar meson masses (all other quantities are essentially unaffected)
$$ F_\pi^* = F_\pi (0) [ 1- R_\pi {\rho \over \rho_0} ] \, , \,
   m_V^*   = m_V   (0) [ 1- R_V   {\rho \over \rho_0} ] \, , \,
   m_\sigma^* = m_\sigma (0) [ 1- R_\sigma {\rho \over \rho_0} ] \eqno(20) $$
where the '*' denotes quantities in the medium and $\rho_0$ is the nuclear
matter density. The range of values for $R_{\pi , V , \sigma}$ is discussed
in ref.42. For simplicity, let me take an universal and equal value, $R_\pi =
R_V = R_\sigma = R$. This is not a direct consequence of the NJL model but
compatible with it. For the sake of the argument I will make here, this
simplification is justified. In ref.42, which is a widely overlooked paper, I
have shown that most of the pv meson--nucleon couplings are very sensitive to
such medium effects, quite in contrast to their strong counterparts. In
particular, the most important pion--nucleon couplings show the following
medium renormalization (for R = 0.2 and at nuclear matter density)
$${G_\pi^* (\rho_0 ) \over G_\pi (0)} = 0.65 \, \, \, \, , \, \, \, \,
  {g_{\pi NN}^* (\rho_0 ) \over g_{\pi NN}(0)} = 0.99    \eqno(21)      $$
and similar results for the vector meson couplings. One can understand this
very different behaviour if one takes a closer look at the expressions for
the various coupling constants. Using the dimensionless variable $x = g F_\pi
r$, with $g$ the universal vector-meson--pion coupling, one notices that the
weak couplings depend one much higher powers of $F_\pi$ than their strong
counterparts and thus are more sensitive to medium modifications. A more
detailed account of this can be found in ref.42.

Finally, let me point out some recent work by Grach and Shmatikov$^{45}$
which concerns yet another mechanism to bring down the value of the pv $\pi
N$ coupling in the medium. The basic idea of thier work is that the
rescattering of emitted pions leads to a strong suppression of $G_\pi$ (the
basic Feynman diagrams are shown in fig.3). Using monopole form factors with
a cut off $\Lambda \simeq 7 M_\pi \simeq 1$ GeV to regulate the diverging
loop integrals, they find
\goodbreak
\bigskip
\midinsert
\vskip 5.5truecm
{\noindent\narrower \it Fig.~3:\quad
Strong pion rescattering in the medium.$^{45}$ The solid, double and dashed
lines denote nucleons, $\Delta$'s and pions, in order. Strong meson--nucleon
vertices are depicted by open circles and weak vertices by the crossed circles.
\smallskip}
\vskip -0.5truecm
\endinsert
$$\eqalign{
G_\pi^{(r)} &= G_\pi \bigl( 1 + {g_{\pi NN}^2 \over 8 \pi^2} I_1 +
 {g_{\pi \Delta N}^2 \over 8 \pi^2} {14 \over 27} I_2 \bigr)   \cr
&= G_\pi ( 1 - 0.76 + 0.01 )                   \cr}      \eqno(22) $$
which leads to
$$ \tilde{F}_\pi^{(r)} = G_\pi^{(r)} g_{\pi NN} / \sqrt{32} \simeq 2.9 \cdot
10^{-7}            \eqno(23)    $$
which is below the bound from the $^{18}$F experiment. This is an interesting
suggestion, but it definitively needs a better treatment (better
regularization procedure) and should also be applied to the other pv
meson--nucleon couplings. Also, one should understand the relation to the
soliton model results discussed before.
\goodbreak
\bigskip
\noindent{OPEN PROBLEMS}
\medskip
Instead of rephrasing what I have said so far, let me just mention the two
salient problems which have to be addressed in the framework of the chiral
soliton model to allow for a deeper understanding of the pv meson--nucleon
interaction regions.
\medskip
\item{$\bullet$}Realistic versions of the three flavor Skyrme model are now
available. They do not indicate a large strange component in proton wave
function. It would be worthwhile and necessary to extend the analysis of the
pv interaction region discussed here. This would also allow to addres such
questions like the strength of the $\phi$--couplings and the relation to the
$\Delta S = 1$ hyperon decay matrix elements. Ultimately, such calculations
will shed some light on the recent developments concerning the possible
enhancement of various weak couplings due to the strange color--singlet
operators.
\medskip
\item{$\bullet$}An old problem is whether the soliton model calculations
should be supplemented by strong interaction enhancement factors or whether
these are already contained in the non--perturbative soliton currents.
This question was to some extent addressed in ref.7, where it was argued that
the inclusion/omission of these factors would at most lead to uncertainties
of the order of 30 per cent, i.e. lead to corrections within  the
accuracy of the model. To my
opinion, this question is not yet settled.  Its resolution will also bring
about the answer to the question of including operators which are not of the
canonical $V_\mu A^\mu$--type.
\bigskip
\goodbreak
\noindent{ACKNOWLEDGEMENTS}
\medskip
It is my pleasure to thank the organizers for their excellent work which made
this meeting truely memorable. I would like to thank  V. Bernard,
D. Driscoll and N. Kaiser for many enjoyable collaborations. I have also
profited from various discussions with B. Desplanques.
\goodbreak
\bigskip
\noindent REFERENCES
\medskip
\nobreak
\item{1.}B. Desplanques, J. F. Donoghue and B. R. Holstein, {\it Ann. Phys.\/}
(NY) {\bf 124} (1980) 449.
\smallskip
\item{2.}T. H. R. Skyrme, {\it Nucl. Phys.\/} {\bf 31} (1962) 556; U.-G.
Mei{\ss}ner and I. Zahed, {\it Adv. Nucl. Phys.\/} {\bf 17} (1986) 143.
\smallskip
\item{3.}U.-G. Mei{\ss}ner, {\it Phys. Rep.\/} {\bf 161} (1988) 213.
\smallskip
\item{4.}N. Kaiser and U.-G. Mei{\ss}ner, {\it Nucl. Phys.\/} {\bf A489}
(1988) 671.
\smallskip
\item{5.}N. Kaiser and U.-G. Mei{\ss}ner, {\it Nucl. Phys.\/} {\bf A499}
(1989) 699.
\smallskip
\item{6.}N. Kaiser and U.-G. Mei{\ss}ner, {\it Nucl. Phys.\/} {\bf A510}
(1990) 759.
\smallskip
\item{7.}M. Shmatikov, Kurchatov Institute of Atomic Energy preprint,
IAE--4636/2 (1988); I. L. Grach and M. Shmatikov,
{\it Z.Phys.\/} {\bf C44} (1989) 393.
\smallskip
\item{8.}E. G. Adelberger {\it et al.\/} {\it Phys. Rev.\/} {\bf C27} (1988)
2833.
\smallskip
\item{9.}J. Dai, M. J. Savage, J. Liu and R. Springer,
{\it Phys. Lett.\/} {\bf B271} (1991)  403.
\smallskip
\item{10.}H. Genz and G. H\"ohler, {\it Phys. Lett.\/} {\bf B61} (1976) 389.
\smallskip
\item{11.}B. R. Holstein, {\it Phys. Rev.\/} {\bf D23} (1981) 1613.
\smallskip
\item{12.}P. Jain, R. Johnson, U.-G. Mei{\ss}ner, N. W. Park and J. Schechter,
{\it Phys. Rev.\/} {\bf D37} (1988) 3252; U.-G. Mei{\ss}ner, N. Kaiser, H.
Weigel and J. Schechter, {\it Phys. Rev.\/} {\bf D39} (1989) 1956, (E) {\bf
D40} (1989) 262.
\smallskip
\item{13.}H. Schnitzer, {\it Phys. Lett.\/} {\bf B139} (1984) 217.
\smallskip
\item{14.}P. Jain {\it et al.,\/} {\it Phys. Rev.\/} {\bf D40} (1989) 855.
\smallskip
\item{15.}M. Gari and J. H. Reid, {\it Phys. Lett.\/} {\bf 53B} (1974) 237.
\smallskip
\item{16.}V. N. Dubovik and S. V. Zenkin, {\it Ann. Phys.\/} (NY) {\bf 172}
(1986) 100.
\smallskip
\item{17.}E. G. Adelberger and W. C. Haxton, {\it Ann. Rev. Nucl. Part.
Sci.\/} {\bf 35} (1985) 501.
\smallskip
\item{18.}U.-G. Mei{\ss}ner, N. Kaiser and W. Weise, {\it Nucl. Phys.\/} {\bf
A466} (1987) 685.
\smallskip
\item{19.}G. N. Epstein, {\it Phys. Lett.\/} {\bf B71} (1977) 267; B. H. J.
McKellar and P. Pick, {\it Nucl. Phys.\/} {\bf B22} (1970) 265.
\smallskip
\item{20.}D. E. Driscoll and G. E. Miller, {\it Phys. Rev.\/} {\bf C39} (1989)
1951; {\bf C40} (1989) 2159.
\smallskip
\item{21.}R. Machleidt, K. Holinde and C. Elster, {\it Phys. Rep.\/} {\bf 149}
(1987)1.
\smallskip
\item{22.}D. E. Driscoll and U.-G. Mei{\ss}ner,
{\it Phys. Rev.\/} {\bf C41} (1990) 1303.
\smallskip
\item{23.}D. E. Nagle {\it et al.\/}, AIP Conference Proceedings No.~51,
p.~224; S. Kistryn {\it et al.\/} {\it Phys. Rev. Lett.\/} {\bf 58} (1987)
1616;
P. D. Eversheim {\it et al.}, {\it Phys. Lett.\/} {\bf B256} (1991) 11.
\smallskip
\item{24.}S. Page, in TRIUMF brown report 89--5, 1989.
\smallskip
\item{25.}W. C. Haxton, {\it Phys. Rev. Lett.\/} {\bf 46} (1981) 698; C.
Bennett, M. M. Lowry and K. Krien, {\it Bull. Am. Phys. Soc.\/} {\bf 25}
(1980) 486.
\smallskip
\item{26.}D. E. Driscoll, PhD Thesis, University of Washington, 1990,
unpublished.
\smallskip
\item{27.}T. Oka, {\it Phys. Rev.\/} {\bf D27}
(1983) 523.
\smallskip
\item{28.}U.-G. Mei{\ss}ner,
{\it Mod. Phys. Lett.\/} {\bf A5} (1990) 1703.
\smallskip
\item{29.}G. B. Feldman, C. A. Crawford, J. Dubach and B. R. Holstein,
{\it Phys. Rev.\/}
{\bf C43} (1991) 863.
\smallskip
\item{30.}D. Eversheim, private communication.
\smallskip
\item{31.}R. R. Silver, W. M. Kloet, L. S. Kisslinger and J. Dubach,
{\it Phys. Rev.\/}
{\bf C40} (1989) 2218;
M. M. Musakhanov and Y. Y. Podgornov,
{\it Yad. Fiz.\/}
{\bf 44} (1986) 709.
\smallskip
\item{32.}N. Kaiser and U.-G. Mei{\ss}ner,
{\it Nucl. Phys.\/} {\bf A506}
(1990) 417.
\smallskip
\item{33.}N. Kaiser and U.-G. Mei{\ss}ner,
{\it Nucl. Phys.\/} {\bf A515}
(1990) 648.
\smallskip
\item{34.}M. Chemtob and B. Desplanques,
{\it Nucl. Phys.\/} {\bf B78}
(1974) 139.
\smallskip
\item{35.}W. C. Haxton, E. M. Henley and M. J. Musolf, {\it Phys. Rev. Lett.\/}
{\bf 63} (1989) 949.
\smallskip
\item{36.}N. Kaiser and U.-G. Mei{\ss}ner, unpublished.
\smallskip
\item{37.}G. S. Adkins and C. R. Nappi,
{\it Phys. Lett.\/} {\bf B137}
(1984) 251.
\smallskip
\item{38.}W.-Y. P. Hwang and K. Nigoyi,
{\it Phys. Rev.\/} {\bf D23}
(1981) 1604.
\smallskip
\item{39.}N. Kaiser, U. Vogl and W. Weise,
{\it Nucl. Phys.\/} {\bf A490}
(1988) 602.
\smallskip
\item{40.}D. B. Kaplan and M. J. Savage, preprint UCSD/PTH 92--04, 1992.
\smallskip
\item{41.}U.-G. Mei{\ss}ner,
{\it Phys. Rev. Lett.\/} {\bf 62} (1989) 1013;
{\it Phys. Lett.\/} {\bf B220}
(1989) 1.
\smallskip
\item{42.}U.-G. Mei{\ss}ner,
{\it Nucl. Phys.\/} {\bf A503} (1989) 801.
\smallskip
\item{43.}V. Bernard, U.-G. Mei{\ss}ner and I. Zahed,
{\it Phys. Rev. Lett.\/} {\bf 59} (1987) 966;
V. Bernard and U.-G. Mei{\ss}ner,
{\it Nucl. Phys.\/} {\bf A489} (1988) 647.
\smallskip
\item{44.}V. Bernard and U.-G. Mei{\ss}ner,
{\it Ann. Phys. {\rm (NY)}\/} {\bf 206} (1991) 50.
\smallskip
\item{45.}I. Grach and M. Shmatikov,
{\it Nucl. Phys.\/} {\bf A536} (1992) 509.
\smallskip
\par
\vfill
\end